\documentclass[twocolumn]{aastex631}

\usepackage{ftnxtra}
\usepackage{ulem}

\def\cd{d$^{-1}$\,}
\def\simlt{\ {\raise-.5ex\hbox{$\buildrel<\over\sim$}}\ }
\def\simgt{\ {\raise-.5ex\hbox{$\buildrel>\over\sim$}}\ }

\received{February 9, 2022}
\revised{March 1, 2022}
\accepted{March 6, 2022}
\submitjournal{ApJL}

\shorttitle{Tidally Tilted Pulsations in an {\rm sdB} Star}
\shortauthors{Jayaraman et al.}
\begin{document}
\title{Tidally Tilted Pulsations in HD\,265435, a subdwarf B Star with a Close White Dwarf Companion}
\correspondingauthor{Rahul Jayaraman}
\email{rjayaram@mit.edu}

\author[0000-0002-7778-3117]{Rahul Jayaraman}
\affiliation{MIT Department of Physics and MIT Kavli Institute for Astrophysics and Space Research, Cambridge, MA 02139, USA}

\author[0000-0001-7756-1568]{Gerald Handler}
\affiliation{Nicolaus Copernicus Astronomical Center of the Polish Academy of Sciences, Bartycka 18, 00-716 Warsaw, Poland}

\author[0000-0003-3182-5569]{Saul A. Rappaport}
\affiliation{MIT Department of Physics and MIT Kavli Institute for Astrophysics and Space Research, Cambridge, MA 02139, USA}

\author[0000-0002-4544-0750]{Jim Fuller}
\affiliation{TAPIR, Walter Burke Institute for Theoretical Physics, Mailcode 350-17, Caltech, Pasadena, CA 91125, USA}

\author[0000-0002-1015-3268]{Donald W. Kurtz}
\affiliation{Centre for Space Research, Physics Department, North West University, Mahikeng 2735, South Africa}
\affiliation{Jeremiah Horrocks Institute, University of Central Lancashire, Preston PR1 2HE, UK}

\author[0000-0002-6018-6180]{St\'ephane Charpinet}
\affiliation{Institut de Recherche en Astrophysique et Plan\'etologie, CNRS, Universit\'e de Toulouse, CNES, 14 Avenue Edouard Belin, 31400, Toulouse, France}

\author[0000-0003-2058-6662]{George R. Ricker}
\affiliation{MIT Department of Physics and MIT Kavli Institute for Astrophysics and Space Research, Cambridge, MA 02139, USA} % as my nominal advisor, he must be included

% Abstract of the paper
\begin{abstract}
Tidally tilted pulsators (TTPs) are an intriguing new class of oscillating stars in binary systems; in such stars, the pulsation axis coincides with the line of apsides, or semi-major axis, of the binary. All three TTPs discovered so far have been $\delta$~Scuti stars. In this Letter, we report the first conclusive discovery of tidally tilted pulsations in a subdwarf B (sdB) star. HD\,265435 is an sdB--white dwarf binary with a 1.65-hr period that has been identified and characterized as the nearest potential Type Ia supernova progenitor. Using TESS 20-s cadence data from Sectors 44 and 45, we show that the pulsation axis of the sdB star has been tidally tilted into the orbital plane and aligned with the tidal axis of the binary. We identify 31 independent pulsation frequencies, 27 of which have between 1 and 7 sidebands separated by the orbital frequency ($\nu_{\rm orb}$), or multiples thereof. Using the observed amplitude and phase variability due to tidal tilting, we assign $\ell$ and $m$ values to most of the observed oscillation modes and use these mode identifications to generate preliminary asteroseismic constraints. Our work significantly expands our understanding of TTPs, as we now know that (i) they can be found in stars other than $\delta$~Scuti pulsators, especially highly-evolved stars that have lost their H-rich envelopes, and (ii) tidally tilted pulsations can be used to probe the interiors of stars in very tight binaries.
\end{abstract}

%%%%%%%%%%%%%%%%% BODY OF PAPER %%%%%%%%%%%%%%%%%%

\section{Introduction}

\subsection{Tidally Tilted Pulsators}

Tidally tilted pulsators (TTPs) are a class of stars in which the pulsation axis of an oscillating star in a close binary system coincides with the line of apsides of the binary, i.e., the semi-major axis (also referred to as the tidal axis), rather than the star's spin axis. The first star thought to have some tidally tilted pulsations was \object{KPD 1930+2752} \citep{2003BaltA..12..139C}, a subdwarf B (sdB) variable star known to be in a short-period binary with a white dwarf companion that is a candidate Type Ia supernova progenitor---not unlike HD\,265435, the subject of this Letter. However, the limitations inherent in ground-based photometric campaigns left this case ambiguous and mostly inconclusive in that respect \citep{2011MNRAS.412..371R}.

Tidally tilted pulsators were conclusively discovered using photometric data from the Transiting Exoplanet Survey Satellite (TESS; \citealt{2015JATIS...1a4003R}) mission. Three TTPs have been robustly identified so far, all using TESS: \object{HD 74423} \citep{2020NatAs...4..684H}, \object{CO Cam} \citep{2020MNRAS.494.5118K}, and \object{TIC 63328020} \citep{2021MNRAS.503..254R}. Other potential tidally tilted pulsator candidates (that are currently described as ``tidally perturbed'') include \object{U Gru} \citep{2019ApJ...883L..26B}, \object{V456 Cyg} \citep{2022arXiv220105359V}, \object{V1031 Ori} (a triple system -- see \citealt{2021PASJ...73..809L}), and \object{VV Ori} \citep{2021MNRAS.501L..65S}. However, further work is required to characterize these four stars' tidally tilted nature. For detailed descriptions and analyses of these intriguing objects, we direct the reader to the recent reviews of \citet{2022arXiv220101722H} and \citet{2020MNRAS.498.5730F}, and the references therein. 

The immediately important property of TTPs is that for binary systems with orbital inclination angles $i \sim 90^\circ$, the observer can view the star through a range of latitudinal angles with respect to the pulsation axis, from $0^\circ$ to $360^\circ$. In turn, this changing view direction with orbital phase enables the observer to better identify the pulsation modes being studied. This does not happen in most single stars or ordinary pulsators in binaries; in these, the observer's view direction remains constant with respect to the pulsation axis. The one exception is encountered in roAp stars (see, e.g., \citealt{1992MNRAS.255..289K}), where the pulsation axis has been tilted with respect to the spin axis by the star's magnetic field.

All three conclusively-identified TTPs are $\delta$~Scuti stars. This naturally raises the question as to whether tidally tilted pulsations can be observed in other types of pulsating stars, such as those that are no longer on the main sequence, or if such pulsations are dependent on some inherent property of $\delta$~Scuti stars. Theory suggests that the latter is unlikely, as the modeling of TTPs in \citet{2020MNRAS.498.5730F} does not in any way rule out tidally tilted pulsations in other kinds of stars. Searches are underway in TESS data to detect these unique systems.

\subsection{sdB Asteroseismology with TESS}

Subdwarf B (sdB) stars are core helium-burning stars with very thin hydrogen envelopes; such objects have been found to exhibit significant chemical peculiarities. These stars are commonly thought to be the stripped cores of red giants, and lie on the Extreme Horizontal Branch of the Hertzsprung-Russell Diagram, with $T_{\rm eff}$ anywhere between 20\,000\,K and 45\,000\,K. For a detailed description of sdB stars and their properties, we point the reader to the review of \citet{heber}.

Many sdB stars are known to pulsate; the first pulsating sdB star was discovered by \citet{1997MNRAS.285..640K}. Since then, over 100 such stars have been discovered \citep{2017MNRAS.466.5020H}. Some sdB stars pulsate with rapid p~mode oscillations, having periods on the order of a few minutes; others pulsate with slower g~mode oscillations, having periods that are on the order of a few hours; and a few stars show a combination of these two. TESS has proven key in the study of rapidly pulsating sdBs, due to the 20-s cadence data mode introduced at the start of the Extended Mission in 2020 July. This ultra-short-cadence mode can probe frequencies up to a Nyquist limit of 2160\,d$^{-1}$, corresponding to periods as short as 40\,s. Consequently, TESS has been at the vanguard of sdB asteroseismology (see, e.g., Section 6 of \citealt{2021FrASS...8...19L}, and references therein). 

\subsection{HD 265435}

HD\,265435 (TIC\,68495594) is an sdB--white dwarf binary that was studied extensively by \citet{2021NatAs...5.1052P} and identified as the closest potential Type Ia supernova progenitor, with a possible thermonuclear detonation occurring in approximately 70 Myr. \citeauthor{2021NatAs...5.1052P} focused primarily on characterizing the properties of the system, including the stellar parameters, the radial velocity of the subdwarf, and the nature of the companion -- which they found to be a white dwarf. 

\citeauthor{2021NatAs...5.1052P} reported that the subdwarf component of HD 265345 is a pulsator with a rich mode spectrum but noted, however, that the Nyquist limit of the 2-min cadence data available to them (360 d$^{-1}$, corresponding to a period of 4\,min) prevented them from conducting a full asteroseismic analysis of the subdwarf component of the binary. This problem is overcome with the latest release of 20-s cadence data from TESS sectors 44 and 45, which show an incredibly rich pulsation spectrum for the sdB star. In this Letter, we use all the available ultra-short-cadence data for this star to characterize its pulsations and study their tidally tilted nature.

\section{Observations}
\subsection{TESS Data}

\begin{figure*}
    \centering
    \includegraphics[width=0.8\textwidth]{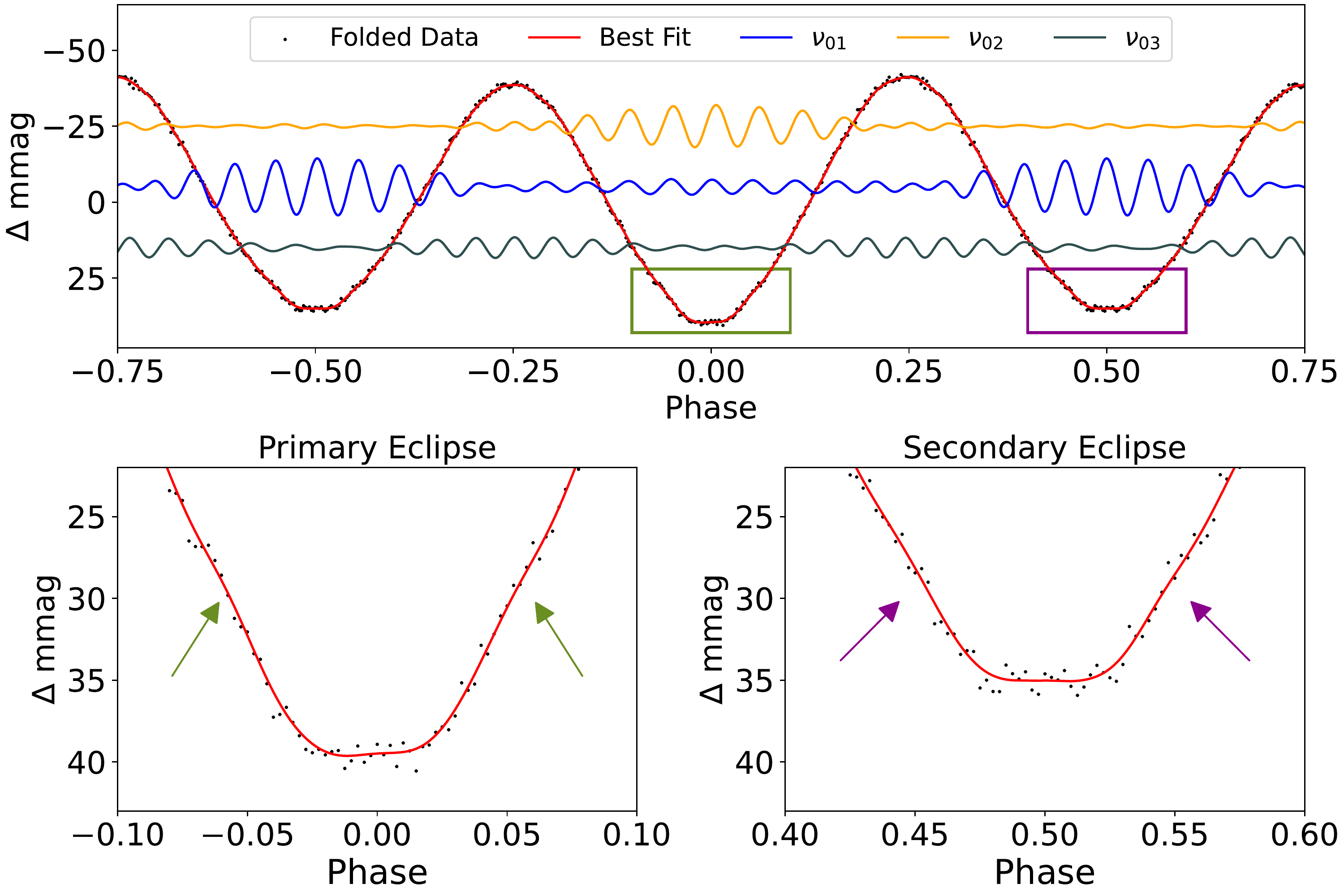}
    \caption{A phase-folded light curve with zooms on possible primary and secondary eclipses. The black points correspond to the actual data, while the red light curve represents a best-fit curve derived from the frequencies removed when prewhitening the orbital harmonics. The boxes in the top panel mark the locations of the zoomed-in plots, where arrows identify the start and end of potential eclipses. The best-fit light curve was smoothed with a Savitsky-Golay filter \citep{1964AnaCh..36.1627S} to remove any short-term variability not arising from the orbital modulations. The blue, yellow, and gray curves in the top panel are reconstructions of the multiplets $\nu_{01}$, $\nu_{02}$ and $\nu_{03}$, with amplitudes equal to their actual values but offset vertically for clarity.}  
    \label{fig:fold_lc}
\end{figure*}

\object{HD 265435} was observed at 2-min cadence in Sector 20 (from 2019 December 24 to 2020 January 21), at 20-s cadence in Sectors 44 and 45 (from 2021 October 12 to 2021 December 2), and at 10-min cadence in the full-frame images during Sector 47 (from 2021 December 30 to 2022 January 28). The shorter-cadence data are available in both SAP (simple aperture photometry) and PDCSAP (presearch data conditioning SAP) forms. Data processing was done using the SPOC pipeline at the NASA Ames Research Center \citep{jenkinsSPOC2016}. We used the 20-s SAP data from both Sectors 44 and 45 to investigate the frequencies of interest. These data span 50.35\,d, and comprise 190\,194 data points, after clipping to remove outliers -- such as those arising from scattered light or cosmic ray strikes on the CCD.\footnote{Data release notes for every TESS sector are available at this link: \url{https://archive.stsci.edu/tess/tess\_drn.html.}}

\begin{figure*}
    \centering
    \includegraphics[width=0.74\textwidth]{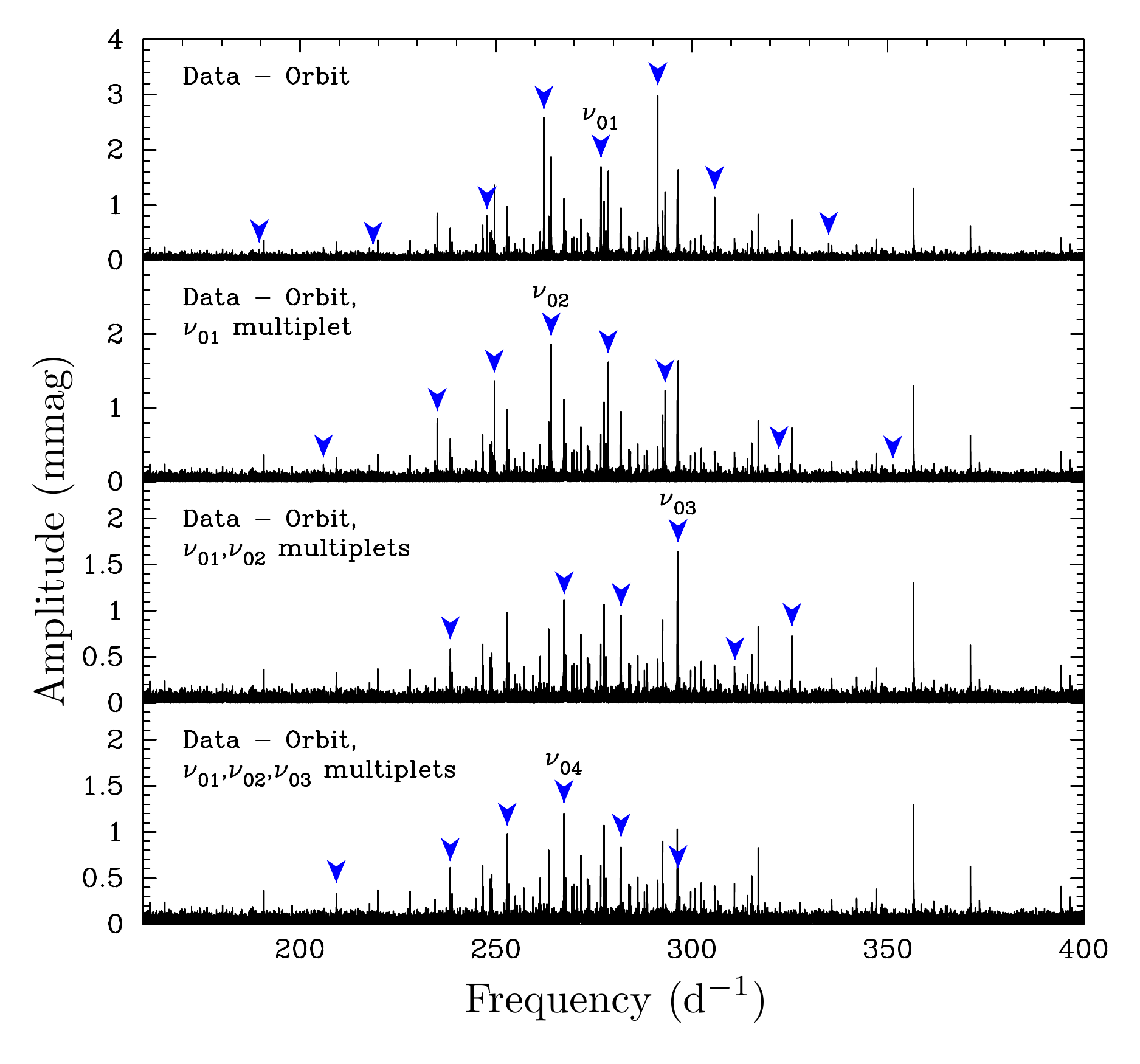}
    \caption{The Fourier amplitude spectrum in the range of the pulsation frequencies, with successive plots (going row-by-row) that highlight the sequential prewhitening of frequency multiplets. The central components of these multiplets are indicated with their identifiers according to Table\,\ref{tab:freqs}, and the blue arrows point at all of the multiplet components.}
    \label{fig:ft_seq_prew}
\end{figure*}

Every sector, we manually review the Fourier spectra of all the 20-s targets observed by TESS to search for particularly high-frequency pulsations in the data. HD\,265435 was flagged for further follow-up due to the combination of: (i) an obvious set of orbital harmonics from a binary light curve (see Figure \ref{fig:fold_lc}), with the first harmonic being the strongest due to ellipsoidal light variations, and (ii) an incredibly rich set of pulsations between the frequencies of 150 and 400 d$^{-1}$ (see Figure \ref{fig:ft_seq_prew}). The presence of both of these features is a rather unusual occurrence in an sdB star. Moreover, the two strongest peaks in the pulsation spectrum appear to be spaced by exactly twice the orbital frequency ($\nu_{\rm orb}$), which prompted us to investigate further. It then became apparent that there are numerous peaks with frequency spacings equal to multiples of $\nu_{\rm orb}$; these were noticed by \citeauthor{2021NatAs...5.1052P} and interpreted as rotationally split modes in the synchronously rotating sdB star.

\section{Pulsational Frequency Analysis}
\label{sec:forb}

A detailed frequency analysis of the TESS data was performed with the {\sc Period04} software \citep{2005CoAst.146...53L}. This package produces amplitude spectra by Fourier analysis and can also perform multi-frequency least-squares sine-wave fitting. It also includes advanced options, such as the calculation of optimal light-curve fits for multiperiodic signals including harmonic, combination, and equally spaced frequencies. The optimal sine-wave fits to the light curve were subtracted from the data, and the residuals were then examined for further periodicities. When deciding to include a given signal in the overall frequency solution, we required its amplitude to exceed the local noise level by a factor of 4.5 (i.e., $S/N > 4.5$). This criterion was relaxed to $S/N>3.5$ in the case of signals at predicted frequencies, i.e., multiplet members or combination frequencies.

We first determined the orbital ephemeris of HD\,265435. In order to calculate this to the highest possible accuracy, we merged the TESS data sets from Sectors 20, 44, and 45 into 6-min bins to suppress any pulsational variations. An initial check showed that the amplitude of the orbital variation was $\sim$10\% smaller in Sector 20 compared to Sector 44, and $\sim$6\% smaller in Sector 45 compared to Sector 44. This is certainly an effect of imperfect correction for the flux of a neighboring star falling into the photometric aperture of our target (for further details, see the Methods section of \citeauthor{2021NatAs...5.1052P}). Thus, we scaled the amplitudes of the Sector 20 and 45 data to match that of the Sector 44 data and determined the ephemeris for the times of the deeper ellipsoidal light minima $T_I$ of the system, corresponding to the orbital phase where the L1 point faces the observer:
\begin{displaymath}
  T_I=2459500.32517(9)+0.068818543(2)E\ \text{(BJD)}.
\end{displaymath}
Here, $E$ is the epoch, i.e., the number of orbital cycles elapsed. We also note that the orbital ephemeris derived by \citeauthor{2021NatAs...5.1052P} contains a typographical error (their BJD$_0$ should actually be 2458909.689955(3)); in contrast to our $T_I$, theirs corresponds to an orbital phase at which the L2 point faces the observer. 

Figure \ref{fig:fold_lc} shows the phase-folded, binned light curve of this system. To calculate the best-fit light curve, we used the amplitudes and phases of the frequencies of the first 100 orbital harmonics. We highlight the flat-bottomed regions near the minima of the ellipsoidal light variations, which hint at the possibility of an eclipsing system. We note that the presence of an eclipse, if our interpretation of this feature is correct, would imply a larger inclination angle than found by \citeauthor{2021NatAs...5.1052P} They estimated an inclination angle between $60^\circ$ and $76^\circ$, whereas eclipses would require an inclination angle of roughly $80^\circ - 85^\circ$. This would, in turn, somewhat lower the mass they determined for the sdB star, and also for the binary system as a whole.

\subsection{The pulsation frequencies}
\label{subsec:puls_freq}

We subtracted the zeropoint in time ($T_0$, as determined above) from the time series, fitted the orbital variation to the data using the nine statistically significant orbital harmonics, and then pursued our frequency analysis with {\sc Period04.} The orbital and pulsation frequency fits were performed simultaneously for the highest possible accuracy and reliability. The top panel in Figure\,\ref{fig:ft_seq_prew} shows the resultant discrete Fourier transform (see, e.g., \citealt{1985MNRAS.213..773K}) covering the frequency range from 160 to 400 d$^{-1}$. The blue arrows mark the most prominent frequency and its sidebands, which are produced by the tidal tilting of the pulsation axis and spaced by the orbital frequency. These groups are called ``multiplets.'' The sequence of three lower panels in Figure\,\ref{fig:ft_seq_prew} shows the sequential prewhitening (described below) of several pulsation frequency multiplets all split by the orbital frequency. This plot illustrates the complexity of the rich pulsation spectrum. 

\begin{figure*}
    \centering
    \includegraphics[width=0.75\linewidth]{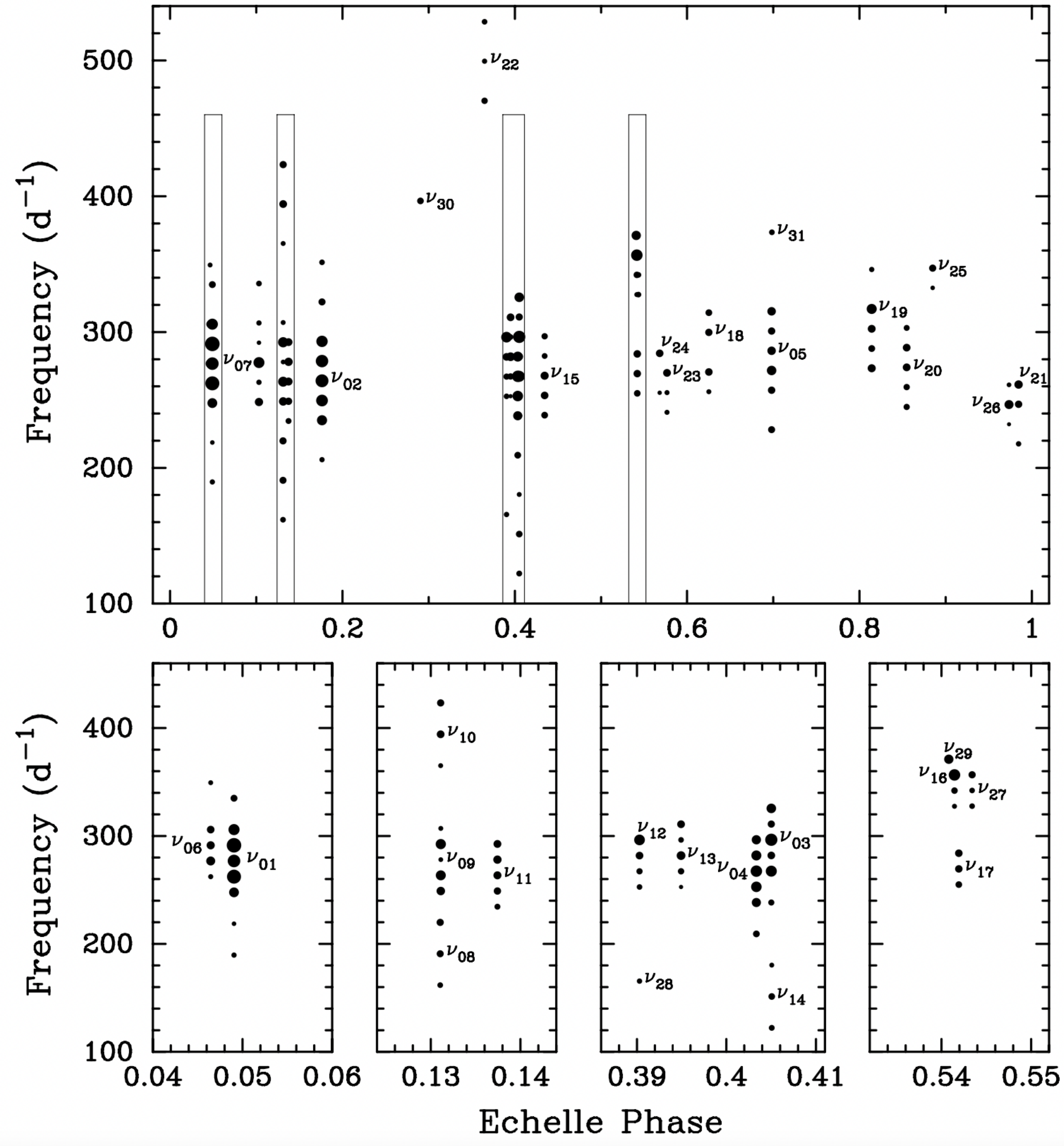}
    \caption{\'Echelle diagram of the pulsation frequencies with respect to the orbital frequency. The size of the filled circles is proportional to the amplitude of that particular constituent of the multiplet. The upper panel shows the full range of frequencies; the lower panels are zooms into regions that are not resolved in the upper panel but contain multiple sequences of frequencies. These regions are denoted with rectangles in the upper panel. The x-axis label "Echelle Phase" is defined as the pulsation frequency modulo the orbital frequency, normalized to the orbital frequency. All observed modes have been labeled.}
    \label{fig:echelle}
\end{figure*}

For each multiplet, we adopt the frequency of the central peak and force the other multiplet frequencies to be separated from it by integer multiples of the orbital frequency. We then do a simultaneous least-squares fit for all the amplitudes and phases of that multiplet and all of the signals determined previously. Then, these signals are all subtracted from the data. This sequential prewhitening process resulted in 31 independent mode frequencies, 90 multiplet components split by integer multiples of the orbital frequency, and seven combination frequencies. The complete frequency solution after simultaneous optimization of all frequencies, amplitudes, and phases for the pulsations is in Table\,\ref{tab:freqs}.  

A convenient way of visualizing multiplets is via an \'echelle diagram, where the frequency of a pulsation is plotted on the vertical axis, and the \'echelle ``phase'' (the pulsation frequency modulo the orbital frequency, normalized to the orbital frequency) is displayed on the horizontal axis. The \'echelle diagram for HD\,265435 is presented in Figure\,\ref{fig:echelle}. Each vertical string of points represents a multiplet, i.e., an independent pulsation frequency split by multiples of the orbital frequency.

As noted previously, \citeauthor{2021NatAs...5.1052P} ascribed the pulsational frequencies of the subdwarf that are visible in the 2-min data to rotationally split multiplets. We argue in the following paragraphs that this is not the case. Rather, the sidelobes of any given pulsation frequency in this star describe the amplitude and phase modulation of that particular mode as a consequence of the time-varying latitudinal viewing angle (with respect to the pulsation axis) over one full orbital cycle.

Rotationally split multiplets have frequencies 
\begin{equation}
\label{eq:ledoux}
\nu_{n,\ell,m} = \nu_{n,\ell} + m(1-C_{n,\ell})\Omega,
\end{equation}
\noindent where $n,\ell,m$ are the radial order, degree, and azimuthal order, respectively; $C_{n,\ell}$ is the ``Ledoux constant''; and $\Omega$ is the rotation frequency. For models of pulsating sdB stars, \citet{2000ApJS..131..223C} find $C_{n,\ell} \simgt 0.02$.

\begin{figure*}
    \centering
    \includegraphics[width=\textwidth]{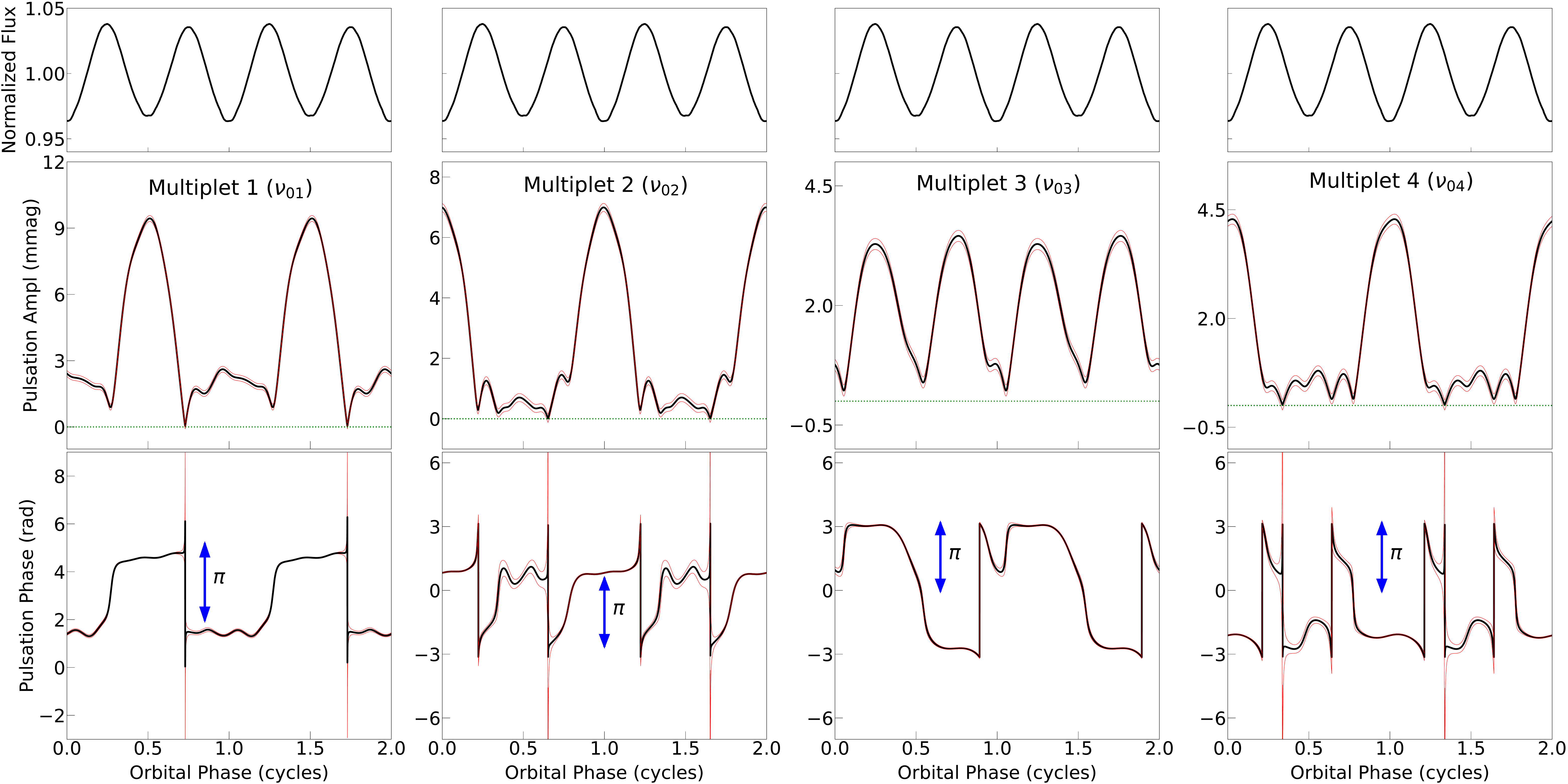}
    \caption{Pulsation phase and amplitudes for multiplets centered on the frequencies $\nu_{\rm 01}$, $\nu_{\rm 02}$, $\nu_{\rm 03}$, and $\nu_{\rm 04}$; these were calculated from the analytical expressions in Equations \ref{eq:analytical_amp} and \ref{eq:analytical_phase}. Red lines represent formal 1-$\sigma$ uncertainties on the values of amplitude and phase; following \citet{1999DSSN...13...28M}, these were calculated as $\sigma_{\rm ampl} = \sqrt{2m/n} \sigma_a$, where the observational rms scatter per single data point $\sigma_a$ =14.61 mmag, $n$ is the number of data points (190,194), and $m$ is the number of detectable peaks in the mode (6-8).  In turn, $\sigma_{\rm ph} = \sigma_{ampl}/a(\Phi_{\rm orb})$, where $a(\Phi_{\rm orb})$ is the amplitude at a given orbital phase $\Phi_{\rm orb}$. Blue arrows on the phase plots, denoting a length of $\pi$, provide a sense of scale. The dotted green lines in the second row of plots denote a pulsation amplitude of zero, i.e., locations where the pulsation phase cannot be meaningfully calculated.}
    \label{fig:amp_phase}
\end{figure*}

For HD\,265435, rotationally split modes are ruled out, as that would require $C_{n, \ell}$ to precisely approach 0 to within the observational error, which is on the order of one part in $10^5$. Furthermore, nearly all of the modes reach pulsation amplitude maximum at the time the tidal axis crosses the line of sight, or $90^\circ$ from that, as expected for oblique pulsation (for a visual representation, see Figure \ref{fig:amp_phase}). This behavior is not expected for a rotationally split mode, as the location of the pulsation amplitude maximum in that particular case would drift with the orbital phase on a timescale set by $C_{n,\ell}$.

Finally, some of the highest amplitude multiplets contain components out to $4 \nu_{\rm orb}$ and $6 \nu_{\rm orb}$, suggesting that these would correspond to high values of $\ell$ in the rotationally split multiplet interpretation. Such high degree modes have very low visibility due to geometrical cancellation \citep{1977AcA....27..203D} and are therefore not a plausible explanation for the modes observed in HD\,265435. As a result, we are able to conclusively rule out the rotational splitting explanation and can conclude that HD\,265435 is a tidally tilted pulsator.

Now that we have established that the frequency multiplets are spaced by exactly the orbital frequency, with the central frequency usually being the actual pulsation mode frequency, we note that there are, in addition, two other effects that contribute to the orbital sidelobes. 

The first arises because of frequency modulation caused by the Doppler shift of the pulsation frequency with orbital motion  \citep{2012MNRAS.422..738S}. That effect contributes up to a few tenths of a mmag to the first orbital sidelobes of the frequency multiplets, but with phases such that there is no contribution of this effect to the measured pulsation amplitude at any orbital phase. In other terms, this effect does not contribute at all to the pulsation amplitudes plotted in Figure \ref{fig:amp_phase} and modeled in Section \ref{sec:modeling}, since Doppler shifts affect only frequency and not amplitude. When describing pulsations using sinusoids, the frequency variability is equivalent to phase variations; thus, this effect does contribute to the phase variations in Figure \ref{fig:amp_phase} but is utterly undetectable. The semi-major axis of the orbit is only 0.83\,R$_\odot$ (as per \citealt{2021NatAs...5.1052P}), and the orbital motion of the sdB star has a radius from the barycenter of only 0.51\,R$_\odot$, which is just 1.2 light-seconds. With pulsation periods on the order of 250\,s, the phase variation caused by the orbital motion ($\sim 0.03$\,rad) is inconsequential. 

The second additional contribution to the orbital sidelobes of the frequency multiplets comes from changes in the background light due to ellipsoidal light variations. Those are $\sim$75\,mmag peak-to-peak; consequently, for a constant pulsation amplitude, they cause an apparent modulation of the pulsation amplitude by 0.075. That will affect the sidelobes' amplitudes, but only to the amount that the pulsation amplitude modulates by -- for example, 0.75\,mmag peak-to-peak for a 10\,mmag pulsation. That is also not the source of the orbital amplitude variations seen in Figure \ref{fig:amp_phase} and in the models that we can now ascribe to tidally tilted pulsations.

\subsection{Reconstructing the Amplitude-Phase Curve of a Multiplet}

\label{subsec:amp_pha}
For each multiplet, we have a collection of amplitudes \{$a_n$\} and phases \{$\phi_n$\}. The amplitudes and phases of the constituent peaks in the multiplet were found as described in Section \ref{subsec:puls_freq}.  All the information about a given multiplet that is available from the data set is also fully contained in the calculated sets of $a_n$'s and $\phi_n$'s.

The time dependence of the pulsation in a given multiplet $\mathcal{M}(t)$ can be written as follows:
\begin{equation}
\mathcal{M}(t) =\sum_{n={\rm -min}}^{n={\rm max}} a_n \cos(\omega_{\rm osc}t+n  \omega_{\rm orb}t+\phi_{n}),
\end{equation}

\noindent
where $t$ is measured from the time where phase zero has been defined, $\omega_{\rm osc}$ and $\omega_{\rm orb}$ are the oscillation and orbital angular frequencies, respectively, and $n = 0$ defines the best estimate of the central element of the multiplet.  

We can expand the cosine function and remove the terms dependent on $\omega_{\rm osc}$ from the summation to find:
\begin{eqnarray}
\mathcal{M}(t) =\cos (\omega_{\rm osc}t) \sum_{n=-{\rm min}}^{n={\rm max}} a_n \cos(n  \omega_{\rm orb}t+\phi_{n})- \nonumber \\ \nonumber \\
\sin(\omega_{\rm osc}t) \sum_{n=-{\rm min}}^{n={\rm max}} a_n \sin(n \omega_{\rm orb}t+\phi_{n})
\end{eqnarray}

We rewrite this expression with basic trigonometric identities in the following suggestive form, where we recast $\omega_{\rm orb}$ in terms of orbital phase $\Phi_{\rm orb}$ as $\omega_{\rm orb}t = \Phi_{\rm orb}$:
\begin{eqnarray}
\mathcal{M}(t) =\mathcal{A}_{\rm osc}(t, \Phi_{\rm orb}, \{a_n\}, \{\phi_n\}) \cos[\omega_{\rm osc}t + \nonumber \\ 
\Phi_{\rm osc}(t, \Phi_{\rm orb}, \{a_n\}, \{\phi_n\})]
\end{eqnarray}
Here, $\mathcal{A_{\rm osc}}$ and $\Phi_{\rm osc}$ are defined as:
\begin{widetext}
\begin{equation}\label{eq:analytical_amp}
\mathcal{A}_{\rm osc} \equiv \sqrt{\left(\sum_{n=-{\rm min}}^{n={\rm max}} a_n \cos(n \Phi_{\rm orb}+\phi_{n})\right)^2+ \left(\sum_{n=-{\rm min}}^{n={\rm max}} a_n \sin(n \Phi_{\rm orb}+\phi_{n})\right)^2}
\label{eqn:amp}
\end{equation}
\begin{equation}\label{eq:analytical_phase}
\Phi_{\rm osc} \equiv {\rm ArcTan2}\left\{\left(\sum_{n=-{\rm min}}^{n={\max}} a_n \sin(n \Phi_{\rm orb}+\phi_{n})\right) ,\left(\sum_{n=-{\rm min}}^{n={\rm max}} a_n \cos(n  \Phi_{\rm orb}+\phi_{n})\right) \right\}.
\label{eqn:phase}
\end{equation}
\end{widetext}
$\mathcal{A}_{\rm osc}$ and $\Phi_{\rm osc}$ are the amplitude and phase of the multiplet and characterize their dependence on $\Phi_{\rm orb}$, the orbital phase. Note that the amplitude and phase of the multiplet do not depend explicitly on the frequency of either the multiplet or the orbit. We have also used the ArcTan2 function to ensure that the phase of the pulsation is located in the correct Cartesian quadrant.

\subsection{The Multiplet Amplitude and Phase Diagrams}
\label{sec:amp_phase}

We utilize the expression for the multiplet amplitude as a function of orbital phase from Equation~(\ref{eqn:amp}) and for the multiplet phase from Equation~(\ref{eqn:phase}) to analytically reconstruct how the pulsation multiplet varies in amplitude and phase around the orbit. Figure \ref{fig:amp_phase} contains these amplitude-phase plots for the four most prominent multiplets. Multiplets $\nu_{01}$ and $\nu_{02}$ have maximum amplitudes when the observer is viewing the L2 and L1 points, respectively. These are $m = 0$ modes, in which the pulsation amplitudes are highly suppressed at one end or the other of the elongated (i.e., tidally distorted) sdB star. There are phase shifts of $\pi$ at the times of ellipsoidal maxima, as expected for such a mode. Multiplet $\nu_{03}$ has its maxima at each of the ellipsoidal maxima, and is thus inferred to be an $|m| = 2$ mode (see Sect.~\ref{sec:modeling}), with no $\pi$ phase jumps.  Multiplet $\nu_{04}$ is very similar in behavior to that of multiplet $\nu_{02}$. Note that the phases after the apparent discontinuity in these two multiplets' plots are identical (phases $\pi$ and $-\pi$ are identical), so this ``jump'' does not represent anything meaningful, unlike the $\pi$ phase shift (discussed previously) that is observed in the phases of multiplets $\nu_{\rm 01}$ and $\nu_{\rm 02}$.

\section{Asteroseismic Modeling}
\label{sec:modeling}

\begin{figure*}
    \centering
    \includegraphics[width=\textwidth]{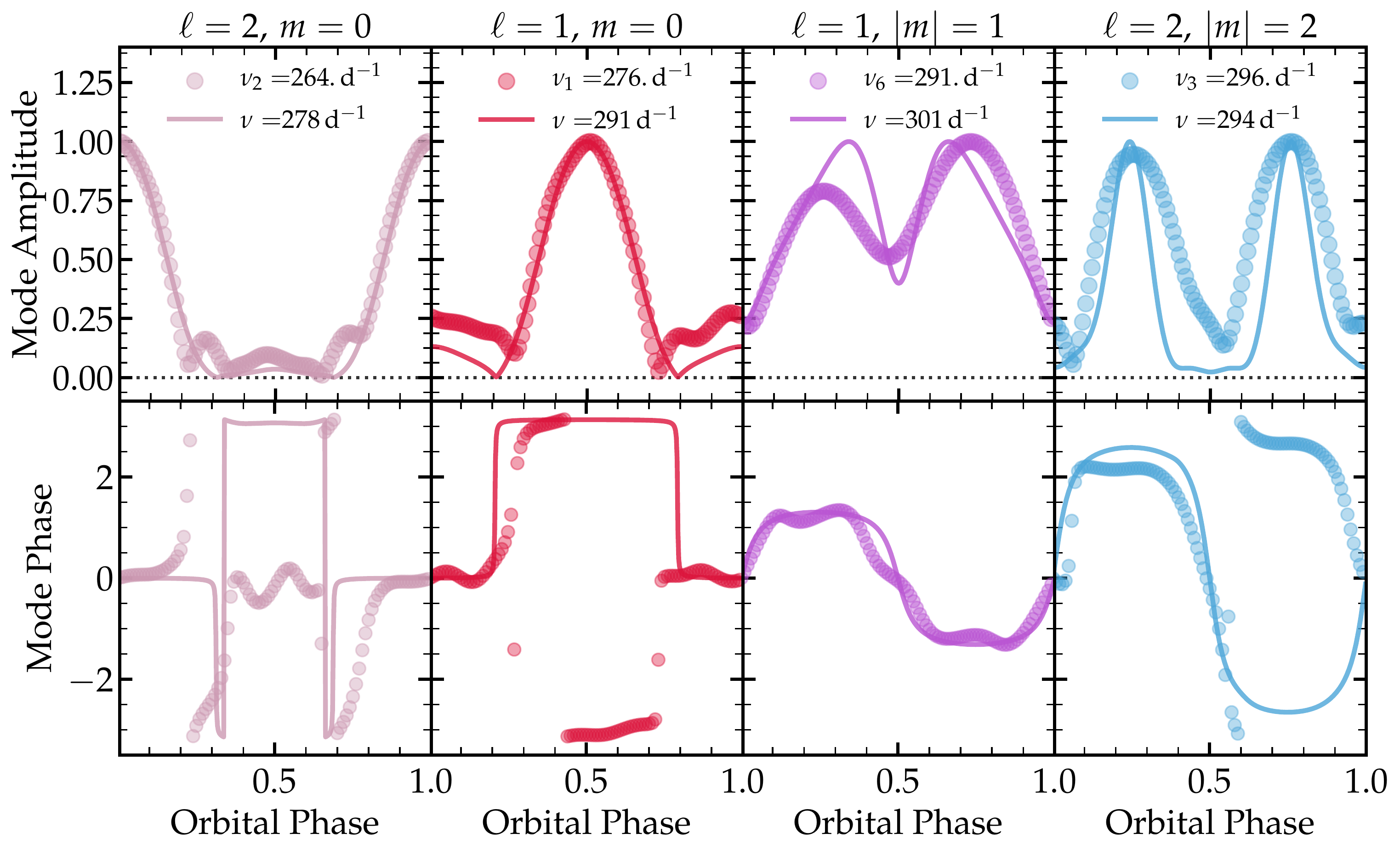}
    \caption{Comparing observed modes (circles) and modeled modes (lines) for a subset of the pulsations of TIC\,68495594. Top panels compare amplitudes (normalized to maximum), while bottom panels compare mode phases, as a function of orbital phase. Columns are labeled by the $\ell$ and $m$ value most likely to correspond to the observed mode. The modeled modes are selected by choosing modes with the appropriate value of $m$, frequencies near $\sim \! 290 \, {\rm d}^{-1}$, and with luminosity fluctuations larger than surrounding modes (such that the mode is not strongly trapped in the core, and is not dominated by high-$\ell$ components). Note that phase $-\pi$ and $\pi$ are identical, so the match with the phases of the $\nu_1$ and $\nu_3$ modes is better than it appears.}
    \label{fig:models}
\end{figure*}

To model the tidally tilted pulsations of HD\,265435, we follow the same procedure described in \cite{2020MNRAS.498.5730F}. We (i) construct a stellar model, (ii) compute non-adiabatic pulsation modes for the spherical star, (iii) calculate tidal coupling coefficients and solve for the new set of tidally coupled modes, and finally (iv) predict the amplitude and phase variation of the calculated modes as a function of orbital phase.

Our sdB models are made with the \texttt{MESA} stellar evolution code \citep{paxton:11,paxton:13,paxton:15,paxton:18,paxton:19}. First, we evolve a $3.2 \, {\rm M}_\odot$ star up the red giant branch and then strip the hydrogen envelope until only the $0.5 \, {\rm M}_\odot$ core remains, which still contains $\sim 0.02 \, {\rm M}_\odot$ of hydrogen. We then enable atomic diffusion (but not radiative levitation) and evolve the star through the core helium-burning phase until it expands slightly to $R\simeq 0.21 \, {\rm R}_\odot$ and $T_{\rm eff} \simeq 34,000 \, {\rm K}$. At this point, we compute oscillation modes using \texttt{GYRE} \citep{townsend:13,goldstein:20}, including modes with $0 \leq \ell \leq 10$ and frequencies ranging from $2 \, f_{\rm dyn} \lesssim f \lesssim 8 f_{\rm dyn}$. Our stellar model has $f_{\rm dyn} = \omega_{\rm dyn}/(2 \pi) \simeq 65 \, {\rm d}^{-1}$, so this frequency range corresponds roughly to the observed range of frequencies shown in Figure \ref{fig:ft_seq_prew}.

In our stellar model, the frequency range where most of the observed modes cluster (260-300$\, {\rm d}^{-1}$) corresponds to the first overtone ($n=2$) $\ell=0$ and $\ell=1$ acoustic modes. For slightly more massive or compact models, the observed frequencies would correspond to the fundamental ($n=1$) $\ell=0$ and $\ell=1$ acoustic modes. Our stellar model has just finished helium burning, so it also contains a radiative core that allows for a fairly dense spectrum of g modes at these frequencies. %The order of the g modes increases for higher values of $\ell$ and for more evolved sdB stars with larger Brunt-V\"ais\"al\"a frequencies in their cores.
For models with a convective helium-burning core, the observed frequencies correspond to low-order gravity modes ($n_g \! \sim \! 1$ for $\ell=1$) trapped just outside the convective core. In either case, the observed modes with non-radial components may be mixed modes. 

Following \citet{2020MNRAS.498.5730F}, we then compute the eigenfunctions of modes aligned with the tidal axis for azimuthal numbers $m=0$, $|m|=1$, and $|m|=2$. By integrating the surface flux perturbations of these modes over the observable hemisphere, we then calculate the mode amplitudes and phases over the orbital cycle. We assume an orbital inclination of $i = 75^\circ$. Figure \ref{fig:models} shows predictions for the amplitude and phase variation of several example modes during the orbital cycle, in comparison to a few observed modes. We find many different types of tidally tilted and tidally trapped pulsations.

For $m=0$ modes, we find examples of modes that are strongly trapped on the ${\rm L}_1$ side of the star (similar to the observed mode $\nu_{02}$, shown in the first panel of Figure \ref{fig:models}), whose amplitudes peak near orbital phase 0. We also find modes that are strongly trapped on the ${\rm L}_2$ side (similar to the observed mode $\nu_{01}$, shown in the second panel Figure \ref{fig:models}), whose amplitudes peak near orbital phase 0.5. There are other axisymmetric modes that are not trapped in either hemisphere (similar to the observed mode $\nu_{07}$) whose amplitudes peak at both orbital phase 0 and 0.5. Axisymmetric modes typically have phase shifts of $\sim \! 0$ or $\sim \! \pi$ over the orbit depending on whether their eigenfunctions are dominated by $\ell = 0$ or $\ell=2$ (like $\nu_{02}$) or $\ell=1$ (like $\nu_{01}$). Note that the modeled mode shown in the left panel of Figure \ref{fig:models} has a phase shift different by $\pi$ compared to the observed mode. This phase shift depends on the relative amplitude of the $\ell=0/2$ and $\ell=1$ components and is difficult to robustly predict, but is unimportant to the tidal trapping phenomenon.

Non-axisymmetric tidally aligned modes have small amplitudes near orbital phases 0 and 0.5 due to geometric cancellation. Instead, their amplitudes typically peak near orbital phases 0.25 and 0.75. Modes dominated by $|m|=1$ and $\ell=1$ will have phase shifts of $\sim \pi$ between these two maxima (like the observed mode $\nu_{06}$), while modes dominated by $|m|=2$ and $\ell=2$ will have phase shifts close to 0 (like the observed mode $\nu_{03}$). Our models find modes with similar variations in amplitude and phase (right two panels of Figure \ref{fig:models}), allowing for tentative mode identifications. Other modes like $\nu_{22}$ (which peaks near orbital phases 0.125, 0.375, 0.625, and 0.875; see Figure \ref{fig:nu_22}) can be ascribed to modes dominated by tidally aligned $|m|=1$, $\ell=2$ modes. Table \ref{tab:frequencies} lists tentative mode assignments for each observed mode, based on the amplitude/frequency modulation described above.

We note that asymmetric mode amplitudes (e.g., the differing maxima in the $\nu_6$ mode shown in Figure \ref{fig:models}) are present for a few modes. This behavior is likely caused by mode mixing between different values of $|m|$ due to the non-axisymmetric components of the Coriolis and centrifugal forces, as discussed in \cite{fuller:20}, which is not included in our models. Tidal distortion and coupling can also shift mode frequencies away from those of a spherical model, complicating asteroseismic analyses. We find that low-order ($n=1$ and $n=2$) $\ell=0$ modes have frequencies decreased by about 10\% relative to the unperturbed model, but these shifts may vary in other models. In general, it is reasonable to expect frequency shifts comparable to the tidal distortion amplitude $\Delta R/R$.

A tentative finding of our pulsation analysis is that the sdB star must be near or past the end of core helium burning. Here, we have identified 14 separate $m=0$ modes in the frequency range 260-300$\, {\rm d}^{-1}$, whereas only about three acoustic modes (the axisymmetric $n=1$ or $n=2$ modes for $\ell=0, \, 1, \, 2$) would be expected. However, models near or just past the end of core helium burning contain many g modes in the observed frequency range, which may explain the large number of observable $m=0$ modes in such a small frequency range. This conclusion is also supported by the modes $\nu_{14} = 151.2\, {\rm d}^{-1}$, $\nu_{08}=190.8\, {\rm d}^{-1}$, $\nu_{03} = 296.5\, {\rm d}^{-1}$, $\nu_{10} = 394.2\, {\rm d}^{-1}$, all of which have amplitude/phase variations consistent with $\ell=2$, $|m|=2$ modes. The lowest frequency modes $\nu_{14}$ and $\nu_{08}$ have frequencies too small to be acoustic modes, and must be predominantly g modes. Assuming they are dominated by $\ell=2$ components, only evolved sdB models have a g mode frequency spectrum dense enough to produce those modes, further favoring models that place the sdB star near the point of helium depletion. Propagation diagrams for models both before and after helium depletion are shown in Figure \ref{fig:propagation}.

%Our models demonstrate that the pulsations of TIC 68495594 can be mostly explained via tidally coupled modes aligned with the tidal axis.
% One issue with our models is that none of the pulsations are predicted to be excited, which is a problem common to many types of pulsators. It is possible that radiative levitation, which was not included in our models, increases iron group element abundances near the mode excitation region, a key feature for exciting low-order acoustic modes in blue large-amplitude pulsators (BLAPs; \citealt{byrne:20}), which have nearly the same temperatures and surface gravities as HD\,265435. A detailed asteroseismic analysis of these pulsations and their excitation requires a separate investigation.

Our models do not include radiative levitation, which is irrelevant for the present analysis, and, not surprisingly, none of the computed pulsation modes are found to be excited (see, e.g., \citealt{1996ApJ...471L.103C, 1997ApJ...483L.123C}). In fact, a similar result is found for Blue Large Amplitude Pulsators (BLAPs), which have nearly the same temperatures and surface gravities as HD\,265435; their pulsations are driven by a similar mechanism (see, e.g., \citealt{byrne:20}). In the case of BLAPs, radiative levitation can enhance iron group element abundances near the mode excitation region. However, we stress that the $T_{\rm eff}$ and surface gravity of HD\,265435 provided by \citeauthor{2021NatAs...5.1052P} place it at the center of the theoretical instability region where p~modes are predicted to be excited in sdB stars, and the observed frequency range is also consistent with that expected from theory \citep{2001PASP..113..775C}. A detailed asteroseismic analysis of these pulsations and their excitation requires a separate investigation.

We conclude by emphasizing that most of our mode identifications are enabled by the tidal tilting phenomenon, which causes phase and amplitude variation that allows us to determine the dominant $\ell$ and $m$ components of the modes. These mode identifications, in turn, allow for more detailed asteroseismic analyses than would be possible without the tidal tilting phenomenon.

\section{Summary and Conclusions}

In this work, we have found that the rich pulsation spectrum of the sdB-white dwarf binary HD\,265435 (TIC\,68495594) contains 27 frequency multiplets between $160-400$\,d$^{-1}$ that are split by the orbital frequency ($\nu_{\rm orb} = 14.5$\,d$^{-1}$).  We have conclusively shown that these multiplet splittings are due to the fact that the sdB star's pulsation axis has been tidally tilted into the orbital plane and is roughly aligned with the binary's tidal axis. Throughout one orbital cycle, the observer views the sdB star over a wide range of latitudinal angles with respect to the tidal axis, causing apparent periodic amplitude and phase shifts of the pulsations with the orbital phase that yield the observed multiplet splittings.  

These amplitude and phase shifts enable us to directly infer the nature of the observed pulsation modes, in many cases enabling us to determine the dominant $\ell$ and $|m|$ component of the mode. In turn, this has led us to a more robust understanding of this star's evolutionary state, finding it is likely near or just past the end of core helium burning. Finally, we have also demonstrated that tidal tilting is possible in both highly evolved stars whose whole envelope has been stripped, as well as in very compact binaries, such as those with $P_{\rm orb} \lesssim 100$ minutes. This should further motivate the search for tidally tilted pulsations in a wider range of binaries.

The TESS data promise to be a continuing source of such discoveries as the satellite continues to survey the sky. Moreover, the combination of upcoming observations and archival data from prior sectors will be beneficial in determining both orbital and pulsational frequencies more precisely. Finally, the 200-s full-frame image cadence in the upcoming Extended Mission 2 (which will enable us to probe up to a Nyquist limit of 231 d$^{-1}$, corresponding to a period of just over 6 min) will enable us to discover even more such stars without the need to request targeted observations at a shorter cadence. 

\section*{Acknowledgments}
G.\,H.~acknowledges support by the Polish NCN grant 2015/18/A/ST9/00578. S.C. acknowledges support from the Agence Nationale de la Recherche (ANR, France) under grant No. ANR-17-CE31-0018, funding the INSIDE project, and financial support from the Centre National d’\'Etudes Spatiales (CNES, France).

This paper includes data collected by the TESS mission. Funding for the TESS mission is provided by the NASA Science Mission Directorate. Resources supporting this work were provided by the NASA High-End Computing (HEC) Program through the NASA Advanced Supercomputing (NAS) Division at Ames Research Center to produce the SPOC data products.

Code and inlists used for our MESA analysis are available on Zenodo, at this\,\dataset[link]{https://zenodo.org/record/6250916\#.Yhf-Qu7MLS5}.

\facilities{TESS}

\software{SPOC \citep{jenkinsSPOC2016}, \texttt{astropy} \citep{astropy:2013, astropy:2018}, \texttt{numpy} \citep{harris2020array}, \texttt{matplotlib} \citep{Hunter:2007}, \texttt{scipy} \citep{2020SciPy-NMeth}, \texttt{pandas} \citep{reback2020pandas, mckinney-proc-scipy-2010}, {\sc{Period04}} \citep{2005CoAst.146...53L}, \texttt{MESA} \citep{paxton:11,paxton:13,paxton:15,paxton:18,paxton:19}, \texttt{GYRE} \citep{townsend:13,goldstein:20}}
\bibliography{ttp-sdb,CoreRotBib}{}
\bibliographystyle{aasjournal}

%%%%%%%%%%%%%%%%%%%%%%%%%%%%%%%%%%%%%%%%%%%%%%
% Don't change these lines

\section*{Appendix}
\startlongtable
\begin{deluxetable}{lcccr}
\tablecaption{Multifrequency solution for the TESS 20-s cadence photometry of HD\,265435. Error estimates for the independent frequencies and pulsation phases at $T_0$ are given in parentheses in units of the last two significant digits; the errors on the amplitudes are $\pm0.05$\,mmag. Modes marked as ``Unclear'' do not have enough information to make a conclusive determination as to their $\ell$ and $m$ values.
\label{tab:frequencies}}
\tablehead{
\colhead{ID} & \colhead{Freq.} & \colhead{Ampl.} & \colhead{Phase} & \colhead{Dominant $\ell$ and $m$} \\
\colhead{} & \colhead{(\cd)} & \colhead{(mmag)} & \colhead{(rad)} & \colhead{}
}
%\colnumbers
\startdata
$\nu_{01}$$-$$6\nu_{orb}$ & 189.6155(23) & 0.22 & 0.40(21) \\
$\nu_{01}$$-$$4\nu_{orb}$ & 218.6775(28) & 0.18 & 2.92(26) \\
$\nu_{01}$$-$$2\nu_{orb}$ & 247.7394(06) & 0.83 & $-$0.09(06) \\
$\nu_{01}$$-$$\nu_{orb}$ & 262.2704(02) & 2.74 & 3.00(02) \\
$\nu_{01}$ & 276.8014(03) & 1.82 & $-$0.18(03) & $\ell = 1, \, m=0$ \\
$\nu_{01}$+$\nu_{orb}$ & 291.3324(02) & 3.17 & 3.02(01) \\
$\nu_{01}$+2$\nu_{orb}$ & 305.8634(04) & 1.19 & $-$0.01(04) \\
$\nu_{01}$+4$\nu_{orb}$ & 334.9253(14) & 0.36 & $-$2.95(13) \\
$\nu_{02}$$-$$4\nu_{orb}$ & 205.9939(23) & 0.23 & $-$0.47(21) \\
$\nu_{02}$$-$$2\nu_{orb}$ & 235.0559(06) & 0.91 & 2.47(05) \\
$\nu_{02}$$-$$\nu_{orb}$ & 249.5869(04) & 1.43 & 2.42(03) \\
$\nu_{02}$ & 264.1179(03) & 1.98 & 2.32(02) & $\ell=2, \, m=0$ \\
$\nu_{02}$+$\nu_{orb}$ & 278.6488(03) & 1.74 & 2.40(03) \\
$\nu_{02}$+2$\nu_{orb}$ & 293.1798(04) & 1.29 & 2.50(04) \\
$\nu_{02}$+4$\nu_{orb}$ & 322.2418(14) & 0.38 & $-$0.47(13) \\
$\nu_{02}$+6$\nu_{orb}$ & 351.3038(21) & 0.25 & 2.67(19) \\
$\nu_{03}$$-$$4\nu_{orb}$ & 238.3809(18) & 0.29 & 0.21(17) \\
$\nu_{03}$$-$$2\nu_{orb}$ & 267.4429(04) & 1.17 & 0.31(04) \\
$\nu_{03}$$-$$\nu_{orb}$ & 281.9739(12) & 0.45 & $-$3.14(11) \\
$\nu_{03}$ & 296.5049(03) & 1.64 & $-$3.11(03) & $\ell=2, \, |m|=2$ \\
$\nu_{03}$+$\nu_{orb}$ & 311.0359(13) & 0.39 & 0.14(12) \\
$\nu_{03}$+2$\nu_{orb}$ & 325.5668(07) & 0.77 & 0.27(06) \\
$\nu_{04}$$-$$4\nu_{orb}$ & 209.2949(15) & 0.34 & 0.77(14) \\
$\nu_{04}$$-$$2\nu_{orb}$ & 238.3569(08) & 0.68 & $-$2.00(07) \\
$\nu_{04}$$-$$\nu_{orb}$ & 252.8879(05) & 1.01 & $-$2.14(05) \\
$\nu_{04}$ & 267.4189(04) & 1.31 & $-$2.19(04) & $\ell=2, \, m=0$ \\
$\nu_{04}$+$\nu_{orb}$ & 281.9498(06) & 0.88 & $-$2.23(05) \\
$\nu_{04}$+2$\nu_{orb}$ & 296.4808(07) & 0.7 & $-$2.05(07)\\
$\nu_{05}$$-$$4\nu_{orb}$ & 228.1098(14) & 0.37 & $-$2.68(13) \\
$\nu_{05}$$-$$2\nu_{orb}$ & 257.1718(13) & 0.41 & 0.16(11) \\
$\nu_{05}$$-$$\nu_{orb}$ & 271.7027(07) & 0.75 & $-$3.01(06) \\
$\nu_{05}$ & 286.2337(09) & 0.56 & $-$3.03(08) & $\ell=2, \, m=0$ \\
$\nu_{05}$+$\nu_{orb}$ & 300.7647(13) & 0.41 & 0.09(12) \\
$\nu_{05}$+2$\nu_{orb}$ & 315.2957(09) & 0.57 & $-$2.88(08) \\
$\nu_{06}$$-$$2\nu_{orb}$ & 262.2327(24) & 0.21 & 0.38(22) \\
$\nu_{06}$$-$$\nu_{orb}$ & 276.7637(08) & 0.65 & $-$2.68(07) \\
$\nu_{06}$ & 291.2947(10) & 0.51 & 0.20(09) & $\ell=1, \, |m|=1$ \\
$\nu_{06}$+$\nu_{orb}$ & 305.8256(12) & 0.44 & 0.56(11) \\
$\nu_{06}$+4$\nu_{orb}$ & 349.4186(26) & 0.2 & 0.95(23) \\
$\nu_{07}$$-$$2\nu_{orb}$ & 248.5255(10) & 0.53 & 0.34(09) \\
$\nu_{07}$$-$$\nu_{orb}$ & 263.0565(23) & 0.23 & $-$2.99(21) \\
$\nu_{07}$ & 277.5874(05) & 1.15 & 0.17(04) & $\ell=0, \, m=0$\\
$\nu_{07}$+$\nu_{orb}$ & 292.1184(29) & 0.18 & $-$3.14(27) \\
$\nu_{07}$+2$\nu_{orb}$ & 306.6494(23) & 0.23 & 0.80(21) \\
$\nu_{07}$+4$\nu_{orb}$ & 335.7114(20) & 0.25 & 0.41(19) \\
$\nu_{08}$$-$$2\nu_{orb}$ & 161.7450(20) & 0.25 & $-$3.09(19) \\
$\nu_{08}$ & 190.8069(14) & 0.37 & 0.32(13) & $\ell=2, \, |m|=2$ \\
$\nu_{08}$+2$\nu_{orb}$ & 219.8689(13) & 0.39 & $-$2.84(12) \\
$\nu_{09}$$-$$2\nu_{orb}$ & 248.9320(09) & 0.57 & $-$0.14(08) \\
$\nu_{09}$$-$$\nu_{orb}$ & 263.4630(06) & 0.87 & 2.88(05) \\
$\nu_{09}$ & 277.9940(27) & 0.19 & $-$0.11(24) & $\ell=1, \, |m|=0$ \\
$\nu_{09}$+$\nu_{orb}$ & 292.5249(06) & 0.92 & 3.00(05) \\
$\nu_{09}$+2$\nu_{orb}$ & 307.0559(25) & 0.2 & $-$0.22(23) \\
$\nu_{10}$$-$$2\nu_{orb}$ & 365.1796(25) & 0.2 & 2.93(23) \\
$\nu_{10}$ & 394.2416(11) & 0.46 & 0.31(10) & $\ell=2, \, |m|=2$ \\
$\nu_{10}$+2$\nu_{orb}$ & 423.3036(13) & 0.39 & $-$3.03(12) \\
$\nu_{11}$$-$$2\nu_{orb}$ & 234.4928(19) & 0.27 & $-$2.82(17) \\
$\nu_{11}$$-$$\nu_{orb}$ & 249.0238(13) & 0.41 & $-$2.92(12) \\
$\nu_{11}$ & 263.5548(10) & 0.49 & $-$3.01(10) & $\ell=2, \, m=0$\\
$\nu_{11}$+$\nu_{orb}$ & 278.0858(10) & 0.54 & $-$2.88(09) \\
$\nu_{11}$+2$\nu_{orb}$ & 292.6167(11) & 0.47 & $-$2.56(10) \\
$\nu_{12}$$-$$3\nu_{orb}$ & 252.6979(21) & 0.25 & $-$0.08(19) \\
$\nu_{12}$$-$$2\nu_{orb}$ & 267.2289(18) & 0.3 & $-$0.43(16) \\
$\nu_{12}$$-$$\nu_{orb}$ & 281.7599(12) & 0.45 & $-$0.51(11) \\
$\nu_{12}$ & 296.2908(05) & 1.06 & 2.77(04) & Unclear \\
$\nu_{13}$$-$$2\nu_{orb}$ & 252.7658(31) & 0.17 & $-$1.50(28) \\
$\nu_{13}$$-$$\nu_{orb}$ & 267.2968(17) & 0.31 & 1.55(15) \\
$\nu_{13}$ & 281.8278(09) & 0.61 & $-$1.90(08) & $\ell=2, \, m=0$ \\
$\nu_{13}$+$\nu_{orb}$ & 296.3588(21) & 0.25 & 1.09(19) \\
$\nu_{13}$+2$\nu_{orb}$ & 310.8897(11) & 0.46 & $-$1.59(10) \\
$\nu_{14}$$-$$2\nu_{orb}$ & 122.1339(19) & 0.27 & 0.32(18) \\
$\nu_{14}$ & 151.1959(16) & 0.33 & $-$2.96(14) & $\ell=2, \, |m|=2$ \\
$\nu_{14}$+2$\nu_{orb}$ & 180.2579(26) & 0.2 & 0.24(23) \\
$\nu_{15}$$-$$2\nu_{orb}$ & 238.8100(15) & 0.34 & $-$1.95(14) \\
$\nu_{15}$$-$$\nu_{orb}$ & 253.3410(12) & 0.44 & $-$1.75(11) \\
$\nu_{15}$ & 267.8720(10) & 0.5 & $-$2.02(09) & $\ell=2, \, m=0$ \\
$\nu_{15}$+$\nu_{orb}$ & 282.4030(20) & 0.26 & $-$1.76(18) \\
$\nu_{15}$+2$\nu_{orb}$ & 296.9339(17) & 0.3 & $-$1.80(16) \\
$\nu_{16}$$-$$2\nu_{orb}$ & 327.5495(25) & 0.21 & 2.16(23) \\
$\nu_{16}$$-$$\nu_{orb}$ & 342.0805(16) & 0.32 & 2.27(15) \\
$\nu_{16}$ & 356.6115(04) & 1.33 & 2.90(04) & $\ell=0, \, m=0$ \\
$\nu_{17}$$-$$\nu_{orb}$ & 254.9016(16) & 0.33 & 1.66(14) \\
$\nu_{17}$ & 269.4326(12) & 0.43 & $-$1.66(11) & $\ell=1, \, m=0$ \\
$\nu_{17}$+$\nu_{orb}$ & 283.9636(12) & 0.43 & 1.32(11) \\
$\nu_{18}$$-$$3\nu_{orb}$ & 256.1142(25) & 0.21 & 0.99(23) \\
$\nu_{18}$$-$$2\nu_{orb}$ & 270.6452(12) & 0.44 & $-$2.09(11) \\
$\nu_{18}$ & 299.7071(13) & 0.41 & 1.20(12) & $\ell=2, \, |m|=2$ \\
$\nu_{18}$+$\nu_{orb}$ & 314.2381(15) & 0.34 & $-$2.06(14) \\
$\nu_{19}$$-$$3\nu_{orb}$ & 273.3898(10) & 0.5 & $-$1.75(09) \\
$\nu_{19}$$-$$2\nu_{orb}$ & 287.9207(15) & 0.35 & 1.31(13) \\
$\nu_{19}$$-$$\nu_{orb}$ & 302.4517(11) & 0.47 & $-$1.64(10) \\
$\nu_{19}$ & 316.9827(06) & 0.88 & 1.61(05) & Unclear \\
$\nu_{19}$+2$\nu_{orb}$ & 346.0447(23) & 0.23 & $-$1.26(21) \\
$\nu_{20}$$-$$2\nu_{orb}$ & 244.9160(17) & 0.3 & $-$1.15(16) \\
$\nu_{20}$$-$$\nu_{orb}$ & 259.4469(17) & 0.31 & $-$1.39(15) \\
$\nu_{20}$ & 273.9779(11) & 0.46 & $-$1.46(10) & $\ell=2, \, m=0$ \\
$\nu_{20}$+$\nu_{orb}$ & 288.5089(11) & 0.45 & $-$1.34(10) \\
$\nu_{20}$+2$\nu_{orb}$ & 303.0399(19) & 0.27 & $-$1.27(17) \\
$\nu_{21}$$-$$3\nu_{orb}$ & 217.7425(22) & 0.23 & 3.03(20) \\
$\nu_{21}$$-$$\nu_{orb}$ & 246.8045(13) & 0.41 & $-$0.16(12) \\
$\nu_{21}$ & 261.3355(09) & 0.55 & $-$0.29(09) & Unclear \\
$\nu_{22}$$-$$2\nu_{orb}$ & 470.2932(17) & 0.31 & $-$0.41(15) \\
$\nu_{22}$ & 499.3552(23) & 0.22 & $-$0.53(21) & $\ell=2, \, |m|=1$ \\
$\nu_{22}$+2$\nu_{orb}$ & 528.4172(21) & 0.25 & 2.79(19) \\
$\nu_{23}$$-$$2\nu_{orb}$ & 240.8715(24) & 0.21 & $-$2.41(22) \\
$\nu_{23}$$-$$\nu_{orb}$ & 255.4025(25) & 0.21 & $-$2.56(22) \\
$\nu_{23}$ & 269.9335(11) & 0.47 & $-$2.43(10) & $\ell=0, \, m=0$ \\
$\nu_{24}$$-$$2\nu_{orb}$ & 255.2800(30) & 0.17 & $-$0.98(27) \\
$\nu_{24}$ & 284.3419(12) & 0.44 & $-$1.34(11) & $\ell=1, \, m=0$ \\
$\nu_{25}$$-$$\nu_{orb}$ & 332.5411(29) & 0.18 & $-$0.11(26) \\
$\nu_{25}$ & 347.0721(13) & 0.39 & 3.10(12) & $\ell=0, \, m=0$ \\
$\nu_{26}$$-$$\nu_{orb}$ & 232.1108(33) & 0.16 & $-$0.39(30) \\
$\nu_{26}$ & 246.6418(08) & 0.65 & $-$0.68(07) & $\ell=0, \, m=0$ \\
$\nu_{26}$+$\nu_{orb}$ & 261.1727(27) & 0.19 & $-$0.77(25) \\
$\nu_{27}$$-$$\nu_{orb}$ & 327.5779(21) & 0.25 & 2.00(19) \\
$\nu_{27}$ & 342.1089(21) & 0.25 & $-$1.98(19) & $\ell=1, \, m=0$ \\
$\nu_{27}$+$\nu_{orb}$ & 356.6399(13) & 0.39 & 1.65(12) \\
$\nu_{28}$ & 165.5120(23) & 0.23 & $-$0.12(21) & Unclear \\
$\nu_{29}$ & 371.1332(07) & 0.72 & 2.81(07) & Unclear\\
$\nu_{30}$ & 396.5577(15) & 0.33 & $-$3.12(14) & Unclear\\
$\nu_{31}$ & 373.4229(20) & 0.25 & $-$0.20(19) & Unclear\\
$\nu_{02}$+$\nu_{04}$$-$$\nu_{orb}$ & 517.0057(18) & 0.29 & $-$2.16(16) \\
$\nu_{02}$+$\nu_{04}$+$\nu_{orb}$ & 546.0677(19) & 0.27 & $-$2.69(18) \\
$\nu_{07}$+$\nu_{09}$$-$$\nu_{orb}$ & 541.0504(19) & 0.27 & 2.51(17) \\
$\nu_{07}$+$\nu_{09}$+$\nu_{orb}$ & 570.1124(18) & 0.29 & $-$0.94(17) \\
2$\nu_{09}$+$\nu_{orb}$ & 570.5189(17) & 0.3 & 1.06(16) \\
$\nu_{09}$+$\nu_{21}$+$\nu_{orb}$ & 553.8604(15) & 0.34 & $-$2.01(14) \\
2$\nu_{09}-\nu_{orb}$ & 541.4569(14) & 0.37 & $-$2.19(13) \\
\enddata
\label{tab:freqs}
\end{deluxetable}

\renewcommand\thefigure{A\arabic{figure}}
\setcounter{figure}{0}    

\begin{figure*}
    \centering
    \includegraphics[width=.8\textwidth]{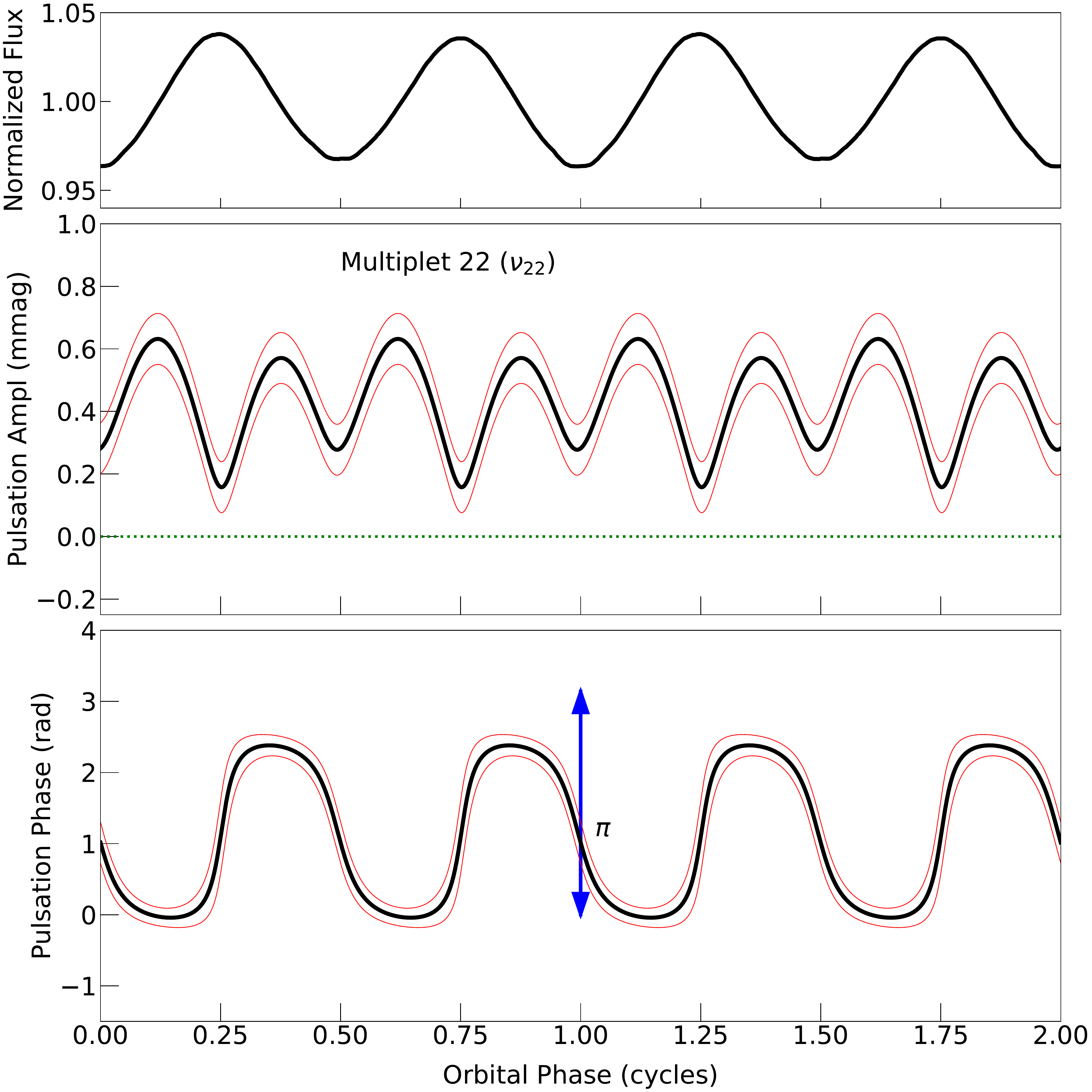}
    \caption{Pulsation amplitude and phase over two orbital cycles for the multiplet centered on $\nu_{\rm 22}$ (similar to Figure \ref{fig:amp_phase}). The pulsation amplitude peaks near orbital phases 0.125, 0.375, 0.625, and 0.875 and has phase ``jumps'' near phases 0.25, 0.5, 0.75, and 1, corresponding to amplitude minima. These can be ascribed to modes dominated by tidally aligned $|m|=1$, $\ell=2$ modes.}
    \label{fig:nu_22}
\end{figure*}

\begin{figure*}
\centering
    \includegraphics[width=.8\textwidth]{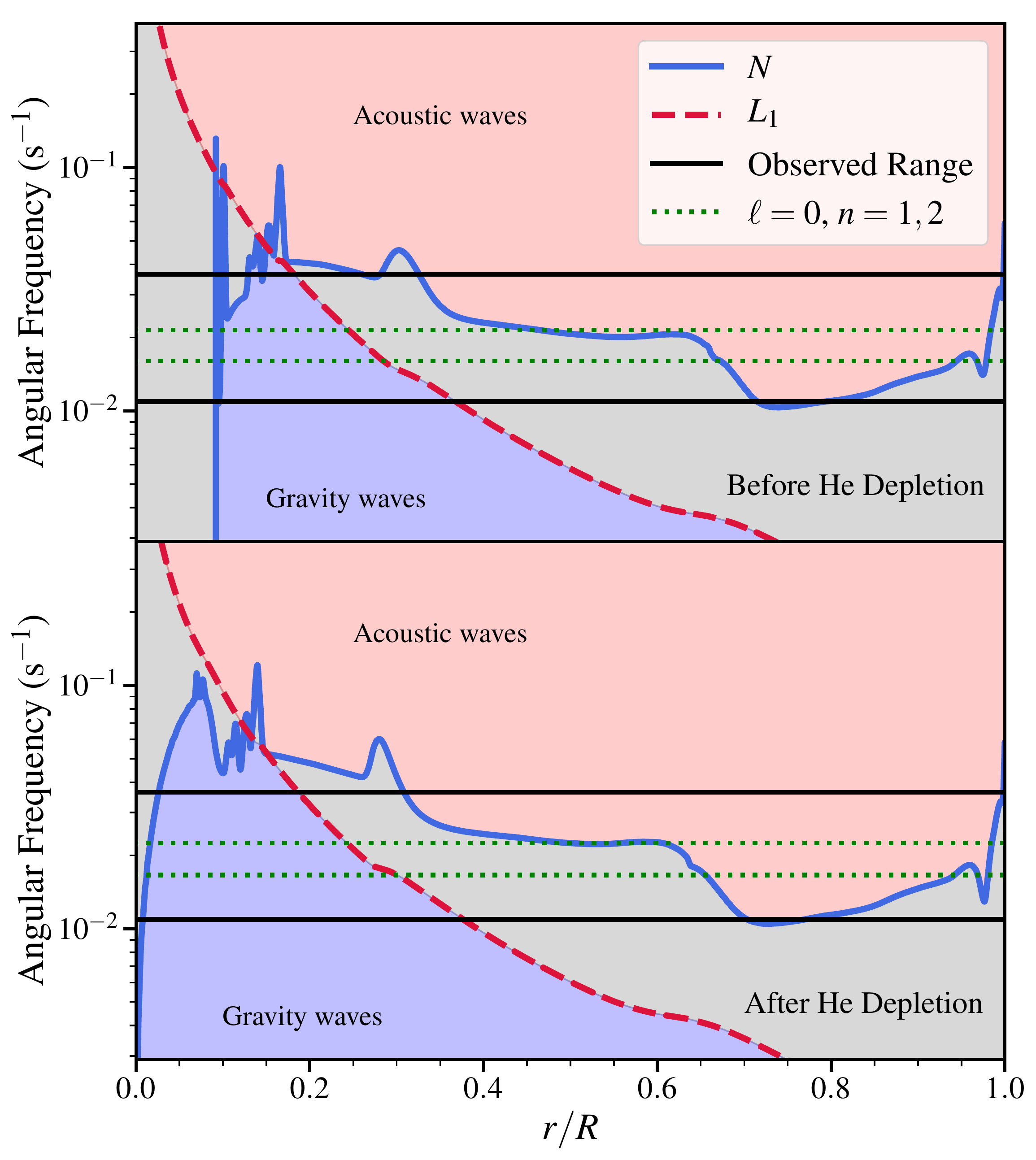}
    \caption{Propagation diagram for our $M=0.5\,M_\odot$ sdB stellar model, showing the Brunt-V\"ais\"al\"a frequency ($N$; blue line), $\ell=1$ Lamb frequency ($L_1$; red dashed line), the lowest order radial modes ($\ell = 0, n = 1,2$; green dotted lines), and the observed pulsation frequency range (black solid lines). The top panel is a model just before core helium depletion, while the bottom panel is just after core helium depletion; the latter is the model used in the main text. Blue shaded regions denote where gravity waves propagate, red shaded regions are where acoustic waves propagate, and grey regions represent evanescent zones.}
    \label{fig:propagation}
\end{figure*}
\label{lastpage}

\end{document}